\documentclass{ws-procs9x6-cpt16}
\begin{document}

\newcommand{\refeq}[1]{(\ref{#1})}
\def\etal {{\it et al.}}

\title{Test of non-Newtonian gravitational force at micrometer range}

\author{Pengshun Luo,$^1$ Jianbo Wang,$^{1}$ Shengguo Guan,$^{1,2}$ Wenjie Wu,$^{1}$ Zhaoyang Tian,$^{1}$ Shanqing Yang,$^{1}$ Chenggang Shao,$^{1}$ Jun Luo,$^{1,3}$ }

\address{$^1$MOE Key Laboratory of Fundamental Quantities Measurement, School of Physics, Huazhong University of Science and Technology, Wuhan 430074, China}

\address{$^2$College of Physics and Communication Electronics, Jiangxi Normal University, Nanchang 330022, China}

\address{$^3$Sun Yat-sen University, Guangzhou 510275, China}

\begin{abstract}
We report an experimental test of non-Newtonian gravitational forces at micrometer range. To experimentally subtract off the Casimir force and the electrostatic force background, differential force measurements were performed by sensing the lateral force between a gold sphere and a density modulated source mass using a soft cantilever. The current sensitivity is limited by the patch electrostatic force, which is further improved by two dimensional (2D) force mapping. The preliminary result sets a model independent constraint on the Yukawa type force at this range.
\end{abstract}

\bodymatter

\section{Introduction}
To connect gravity with the rest of physics, many theoretical frameworks beyond the standard model of particle physics have been proposed, where either hypothetical interactions were predicted to exist by exchanging new bosons or the gravitational force would deviate from the Newton's inverse square law at sub-millimeter range when large extra dimensions is considered.\cite{ADD, Banks1988, Beane1997, Sundrum2004}  Many experiments have been done to search for such forces or deviations (both them are termed as non-Newtonian gravitational forces).\cite{Mohideen2000, Decca2007, Sushkov2011, Decca2016, Stanford, HUST, Washington, Adelberger2003, Adelberger2009, Newman2009} The formalism of the forces could be Yukawa type or power law type, and experimentalists used to parameterize the results using the following potential
\begin{equation}
V(r) = -\frac{Gm_1m_2}{r}(1+\alpha e^{-r/\lambda}).
\label{YukawaPotentail}
\end{equation} 

The strongest constrains on such forces have been obtained by the torsional balance experiments at millimeter or sub-millimeter ranges.\cite{HUST, Washington} The sensitivity is usually below the Newtonian gravitational force which is the dominant background to be considered. At micrometer range, the main challenge for such experiment is the intervening of the strong Casimir force and the electrostatic force background. The current constraints were mostly derived from the precision measurement of the Casimir force,\cite{Mohideen2000, Decca2007, Sushkov2011}  which depend on the reliability of the Casimir force calculation and the evaluation of the patch electrostatic force.\cite{KLIMCHITSKAYA} Therefore, it is important to perform a model independent experiment to set a reliable constraint without subtraction of the calculated forces. 

\section{Experiment}
The idea to subtract off the background force is to perform a differential measurement. In our experiment, a soft cantilever attached with a gold sphere is used to sense the lateral force acting on the sphere by the source mass. The source mass is composed of a density modulated structure with alternative high density (gold) and low density (silicon) materials, which produces a modulated force field. By driving the source mass oscillating under the gold sphere at constant separation, the gold sphere could sense the gravitational force modulation. It should be noted that the Newtonian gravitational force is several orders of magnitude smaller than the experimental sensitivity. 

To subtract off the Casimir force and the electrostatic force, we deposited a layer of gold on top of the density modulation structure, so that the two forces are in principle constant during the source mass oscillation. As the two forces are mainly normal to the source mass surface, we placed the cantilever normal to the surface to let it sensitive to the lateral force. It is critical to make a flat and well conductive surface in this experiment. To achieve this goal, we fabricated the source mass based on a silicon on insulator (SOI) wafer where the silicon dioxide layer serves as a template for the flat surface required.  

The source mass is driven to oscillate with an optimal amplitude so that the possible signal is expected at 8 times of the drive frequency $f_d$. The force sensitivity is around $3\times 10^{-15} N/\sqrt{Hz}$ at the signal frequency. We performed such measurement on a 2D grid parallel to the source mass surface, so that we obtained a 2D image of the force signal at 8$f_d$. 

\section{Preliminary result}
We observe that the 2D image of the force signal depends on the thickness of gold coating and post-annealing process. The reason is believed to be the suppression of the patch electrostatic force by better conductivity between the gold grains. For the source mass with 500 nm gold coating and post-annealing at 150 $^{\circ}$C for 12 hours, we do not observe obvious correlation between the signal and the modulation structure except for random fluctuation. The 2D image is then fit to the Yukawa force by maximum likelihood estimation, which gives a best fit $\alpha$ value and its standard deviation $\delta\alpha$ for every $\lambda$. The constraint at 95$\%$ confidential level is then set by $\alpha + 2\delta \alpha$. The constraint set by this work is about 2 times weaker than the best result obtained from the Casimir experiment.\cite{Sushkov2011} We noted that a much better result was reported recently by the Indiana University--Purdue University Indianapolis group (presented also at this conference).\cite{Decca2016}

\section*{Acknowledgments}
This work was supported by the National Basic Research Program of China under Grant No. 2010CB832802 and the National Natural Science Foundation of China under Contract Nos. 11275076 and 91436212.

\end{document}